# On the Use of Information Retrieval to Automate the Detection of Third-Party Java Library Migration at the Method Level

Hussein Alrubaye*, Mohamed Wiem Mkaouer*, Ali Ouni†
*Software Engineering Department, Rochester Institute of Technology, NY, USA
†ETS Montréal, University of Quebec, Montréal, QC, Canada
{hat6622,mwmvse}@rit.edu, ali.ouni@etsmtl.ca

*Abstract*—The migration process between different third-party libraries is hard, complex and error-prone. Typically, during a library migration, developers need to find methods in the new library that are most adequate in replacing the old methods of the retired library. This process is subjective and time-consuming as developers need to fully understand the documentation of both libraries' Application Programming Interfaces, and find the right matching between their methods, if it exists. In this context, several studies rely on mining existing library migrations to provide developers with by-example approaches for similar scenarios. In this paper, we introduce a novel mining approach that extracts existing instances of library method replacements that are manually performed by developers for a given library migration to automatically generate migration patterns in the method level. Thereafter, our approach combines the mined method-change patterns with method-related lexical similarity to accurately detect mappings between replacing/replaced methods. We conduct a large scale empirical study to evaluate our approach on a benchmark of 57,447 open-source Java projects leading to 9 popular library migrations. Our qualitative results indicate that our approach significantly increases the accuracy of mining method-level mappings by an average accuracy of 12%, as well as increasing the number of discovered method mappings, in comparison with existing state-of-the-art studies. Finally, we provide the community with an open source mining tool along with a dataset of all mined migrations at the method level.

## I. Introduction

Modern software systems rely heavily on third-party library functionality as a mean to save time, reduce implementation costs, and increase their software quality when offering rich, robust and up-to-date features [1], [2], [3]. However, as software systems evolve frequently, the need for better services and more secure, reliable and quality functionalities causes developers to often replace their old libraries with more recent ones. This process of replacing a library with a different one, while preserving the same functionality, is known as *library migration* [4].

The migration between two given libraries consists of a sequence of steps: It starts with retiring the current library by removing all its dependencies from the program, which includes imports and method calls. Developers are then required to find the right replacing method(s) for each removed method belonging to the retired library. Developers are also required to verify whether the newly adopted methods are delivering the same expected functionalities of the retired library's methods. These steps tend to be subjective, time-consuming, and error-prone, as developers need to fully understand both retired and new libraries methods, and be aware of their implementation details. This includes the exploration of their Application Programming Interfaces (API) documentation and the online search for code snippet examples of their methods usage. Moreover, the matching process between the replaced and replacing methods, belonging respectively to the retired and new library, is not straightforward. Even if libraries offer similar services, they may be different in their methods design and documentation. Thus, it would be beneficial to learn from the collective experience based on the manually performed library migrations in the past.

However, the detection of such migrations is challenging. First, there is no systematic way to detect the developer's intention of adopting a library migration. Therefore, its detection may require extensive analysis of the history of code changes while searching for specific replacement patterns between APIs. Furthermore, deciphering the pairing of removed and added methods is complex especially when many of them are co-located in the same code block. In addition, there is no strict rule about the cardinality of the pairs, *i.e.*, one or many methods from the replaced API can be replaced by one or many methods from the new API, which makes its automated detection more challenging.

Several studies have tackled the problem of identifying the pairs of removed and added methods, also known as *mappings*, using Information Retrieval (IR) techniques to detect method change patterns, method signature similarity, and method graph mining [4], [5], [6]. These approaches have provided efficient results when finding 1-to-1 mappings between methods. However, they are mainly challenged when identifying mappings with larger cardinality, *i.e.*, when one or many methods can be replaced with one or many methods, also known as *one-to-many* or *many-to-many* method mappings. Also, when two or more source methods, located in the same code block, are being replaced by two or more target methods, this creates another challenge to distinguish between these interleaved mappings.

This paper builds on the existing studies by leveraging the lexical similarity with the repetitiveness of code changes in software systems in general, and in migrations in particular [7]. We mine the repetitive patterns of method replacements, *i.e.*, mappings in the code. Intuitively, the more a mapping between methods is detected across several code fragments, the more relevant this mapping becomes, *i.e.*, a pattern.

Furthermore, to cope with interleaved mappings *i.e.* mappings occurring in the same code blocks, we identify potential mappings between methods, based on how similar their signatures and API documentation.

We implement our approach in an open-source tool that identifies all method- level migration traces between given libraries throughout a set of representative projects. For a given library, our approach works as follows: (1) it first mines all projects to identify all migration segments, *i.e.*, set of code changes (*e.g.*, commits), in which developers performed code changes related to the given libraries APIs; (2) it extracts all migration fragments *i.e.*, code diffs containing set of removed and added methods; (3) it generates all mappings between all removed and added methods, *i.e.*, each removed method belonging to the retired library will be to one or multiple added methods belonging to the new library.

As an attempt to evaluate our approach, we conduct a large scale empirical study on a benchmark of 57,447 open-source Java projects mined from GitHub. Results show that our approach outperforms three state-of-the-art approaches by achieving an average accuracy improvement of 12%, as well as increasing the number of discovered mappings by 17%. Furthermore, the quantitative analysis of our results indicates that our approach requires less number of code fragments to accurately extract all mappings.

The paper has the following main contributions:

- We introduce a novel approach that increases the accuracy of detecting migration fragments during the library migration process.
- We conduct a large-scale empirical study on 57,447 open-source Java projects while mining 9 popular library migrations. We also conduct a comparative study between our approach with three state-of-art approaches that we adapt for the library migration problem.
- We provide an open-source tool along with the generated migration results as a dataset for the research community to better comprehend how developers practice library migrations[1].

[1]http://migrationlab.net/index.php?cf=icpc2019

The paper is structured as follows: Section II unlocks the terminology that is used throughout the paper. Section III enumerates the studies relevant to our problem. Section IV details the challenges related to extracting existing library migrations through a motivating example. Our approach is detailed in Section V, it also issues an example to illustrate how each of the approaches under comparison generates the mappings at the method level. Section VI shows our experimental methodology in collecting the necessary data for the experiments that are discussed in Section VII. Finally, the conclusion and future work are highlighted in Section VIII.

## II. BACKGROUND

This section presents definitions of the main concepts that are used throughout the paper.

**Migration Rule.** A migration is denoted by a pair of a source (retired) library and a target (replacing) library, *i.e.*, *source* → *target*. For example, *easymock* → *mockito* represent a migration rule where the library easymock[2] is migrated to the new library mockito[3]. Table I depicts the list of migration rules that are mined and studied in this paper.

**Method Mapping.** A migration rule is a set of method mappings between the source and the target library. The mapping between methods is the process of replacing a least one method from the source library by one or multiple methods belonging to the target library. Figure 4(E) shows some examples of mappings.

**Segment.** It constitutes the migration *period*. It is a sequence of one or multiple code changes (*e.g.*, commits), containing each, one or multiple fragments.

**Fragment.** A block of source code that witnesses at least one mapping. It is generated by contrasting the code before and after the migration to only keep the removed (resp., added) methods linked to the source (resp., target) library. For example, Figure 2 depicts three fragments, each fragments contains a set of added/removed methods.

## III. RELATED WORK

Several studies focused on understanding how developers perceive API related method changes. In the context of library updates, many studies have been proposed to capture the needed changes on the client source code applied along with API migration [8], [9], [10], [11]. Most of the existing approaches use textual similarity between the structures and method signatures as a basic technique to identify identical methods between multiple library versions. Similar approaches were tackling the problem of mapping between methods across different languages. The majority of these approaches employed information retrieval and natural language processing

[2]http://easymock.org
[3]https://site.mockito.org

techniques to identify similar method usages in different languages [12], [13].

Another recent study has been conducted by Schäfer et al. [6], by analyzing changes in the method call locations to extract the fragments of added/removed methods. The authors compute associated rules from the fragments before filtering them using the similarity of method signatures. The approach allocates one method to each call. Consequently, such approach favors the *one-to-one* method mapping and ignores the existence of other added (resp., removed) methods in case of *many-to-many* method mapping *i.e.*, replacing one or many methods with one or many methods within the same fragment.

Teyton et al. [5] extended this to support all possible cardinalities of method mappings. They performed the same migration process for a given input migration rule to extract all the fragments, then they applied the Cartesian Product between the two sets of removed and added libraries. This generates all the possible combinations of mappings that may have occurred between the set of source library and target library methods. Then, they calculate the frequency of identical combinations throughout all the studied projects. Finally, they define an acceptance threshold where any combination with a higher frequency than this fixed threshold is considered a legit method mapping.

In contrast with previous studies, this approach was similarity-agnostic since it is robust against libraries variations in their design, naming conventions and vocabulary. On the other hand, it exclusively relies on existing migrations between two given libraries to be able to provide mappings. Lastly, its performance in terms of accuracy depends on the frequency of such migrations across projects, as we will discuss later in Section VII.

Similarly, Hora et al. [14], [15] adapted the approach of Teyton et al. [5] in the context of detecting method mappings between different releases of the same library in order to analyze the evolution of its API. They used association rules and the frequent itemset mining technique on method call changes between two versions of the same method. The proposed approach generated thereafter rules to specify which old call should be replaced with a new call. The study was extended later in [16] to analyze the developers' perception of these tracked API changes.

In this paper, we perform a comparative study between our approach, Teyton et al. [5], and Schäfer et al. [6]. We also adapted the approach of Nguyen et al. [10] in the context of detecting mappings by pairing methods that have a strong similarity in their signatures. In the next section, we provide the challenges of detecting mappings at the method level, we also show how our approach generates the existing method mappings through an illustrative example.

## IV. The Migration Dilemma

The migration process between two different libraries is a hard, error-prone and time-consuming process [1], [2], [17], [18].

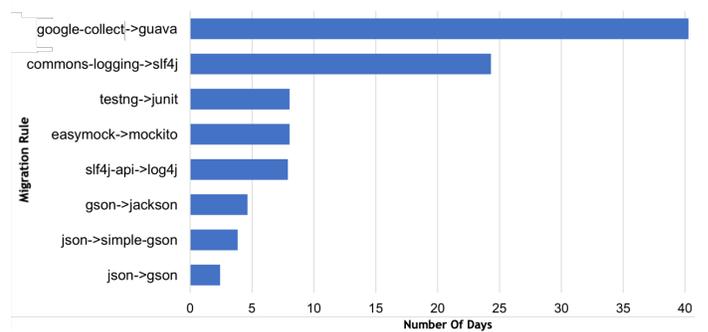

Figure 1: Average time spent by developers to perform the migration between pairs of popular Java libraries.

To showcase the complexity of the process, we measure, based on the data we collected in our experiments, the average time spent to perform migration between different libraries as shown in Figure 1. We approximate the migration time by calculating the difference between timestamps of the first and last commit that contained migrations, as the migration is typically performed in multiple commits [17]. Figure 1 shows that, depending on how complex is the migration, developers typically spend from 2 to 42 days to migrate between libraries.

Furthermore, on an ideal setting, each source library method is replaced with one target library method (*one-to-one*), in each fragment. This makes their detection easier and less error-prone. In practice, due to the differences in libraries design, separation of concerns, and naming conventions, a method may be replaced with more than one method from the target library (*one-to-many*). Furthermore, there exists different co-locations of many added and removed methods within the same source block, which makes the automated identification of individual mappings more complex.

As a motivating example, depicted in Figure 2, we consider three fragments from Github[4] that were extracted as part of the migration from *json* to *gson*. Each fragment contains a replacement scenario that is described as follows:

- **One-to-one mapping.** It is the replacement of one method with another method. In Figure 2-(A), the method *put(key, value)* is replaced by one method, namely *addProperty(key, value)*.
- **One-to-many mapping.** Replacing one method with more than one method. In Figure 2-(B), the method *put(key, value)* has been replaced with two methods, namely *addProperty(key, value)*, and

[4]http://migrationlab.net/redirect.php?cf=icpc2019&p=1

```java
public void addKeyValues(String key, int value) {
    checkIfKeyDescriptionExist(key);
-   keyValues.put(key, value);
+   keyValues.addProperty(key, value);
}
```
(A) One-to-one

```java
public void addKeyValues(String key, Map value) {
    checkIfKeyDescriptionExist(key);
-   keyValues.put(key, value);
+   keyValues.addProperty(key, new Gson().toJson(value));
}
```
(B) One-to-many

```java
-   if (!endpoints.isEmpty()) {
-       node.put("endpoints", endpoints);
+   if (!endpoints.isJsonNull()) {
+       node.add("endpoints", endpoints);
```
(C) Many-to-many

Figure 2: Samples of migration between *json* and *gson*.

*Gson().toJson(value)*. To have valid input for *addProperty* method, the *Map* object needs to be converted into a *json* object, so another converting method was added. Note that, in Figure 2(A), the same method was replaced with only one method since the input *Value* had no mismatch.

- **Many-to-many mapping**. Replacing many methods (two or more) with two methods (two or more). As a real-world example, Figure 2-(C) shows how the two methods *isEmpty()*, and *put(key, value)* have been replaced with two different methods, namely *isJsonNull()* and *add(key, value)*.

- **Multiple correct mappings.** A method from the removed library could be mapped to more than one method from the added library. For example, In Figure 2-(A), the method *put(key, value)* was replaced by *addProperty(key, value)*, while in Figure 2-(C), the same method was replaced by *add(key, value)*. The reason behind the developer's decision could be related to opting a more relative method that requires fewer changes. Otherwise, if the developer decides to replace *put(key, value)* with *addProperty(key, value)*, then another method would likely to be added, namely *getEndpoint()* to get Integer value of *endpoints*, so the added code would be *addProperty("endpoints", endpoints.getEndpoint())*. So there are multiple possible correct mappings for a given method. This scenario indicates how a migration is merely a subjective process, and developers tend to choose simpler *one-to-one* mappings, whenever it is possible to reduce the amount of unnecessary changes. We follow the same intuition in our approach as we opt for mappings with lower cardinality, as much as possible during our mapping generation process.

As shown in the motivating example, a large number of type mappings could be extracted from one single code change, *i.e.*, commit. This is a particularly challenging task to generate accurate and relevant library method mappings to support the library migration process. Indeed, existing state-of-the-art approaches relying basically on lexical similarity achieved a limited accuracy in identifying *one-to-many* or *many-to-many* mappings. To address these limitations, our approach intersects fragments to generate all possible mapping types between methods, then calculates the frequency of each mapping across all fragments. Intuitively, the higher the frequency of a mapping is, the more relevant it is for the migration. Furthermore, we intersect mappings with lower cardinality (*one-to-one*) with those having higher cardinality (*one-to-many* and *many-to-many*) in order to reduce their cardinality. Additionally, we use the similarity between the documentation of API methods as it would provide a rich and meaningful information to reduce the cardinality of *many-to-many* mappings by extracting a *one-to-one* mapping particularly in cases where the combination of the methods documentation exhibit a strong similarity.

V. Substitution Algorithm

In this section, we introduce our approach for generating method mappings for library migration. Figure 3 provides an overview of our approach which consists of four main phases: (1) collection phase, (2) segment and fragment detection phase, (3) mapping generation phase, and (4) validation phase. In the following, we explain each of these phases. For the fragments collection and detection, we reuse dataset of our previous study[17].

*A. Collection Phase*

The collection phase takes as input a list of open-source Java software systems projects. It starts by cloning and checking out all commits for each project. For every commit, we collect its properties, such as commit ID, commit date, developer name, and commit's description. We also keep track of all changes in the project library configuration file, known as Project Object Model (*pom.xml*). All mined projects data is recorded in a database for faster querying when conducting the identification of segments and fragments.

*B. Detection Phase*

The detection phase consists of the identification of (1) segments and (2) fragments.

**Segment Detection:** The purpose of the segments detection, *i.e.*, migration periods, is to locate, for each migration rule its time periods in all projects. As defined in the background section, a segment could be composed of one or many fragment-related commits involved in the migration process. As shown in Figure 3, the segment detection phase starts with checking whether both libraries exist in the list of added/removed project libraries. Using static code analysis, this phase locates the end of the segment by scanning all commits in which all project source files are no longer dependent on the retired library.

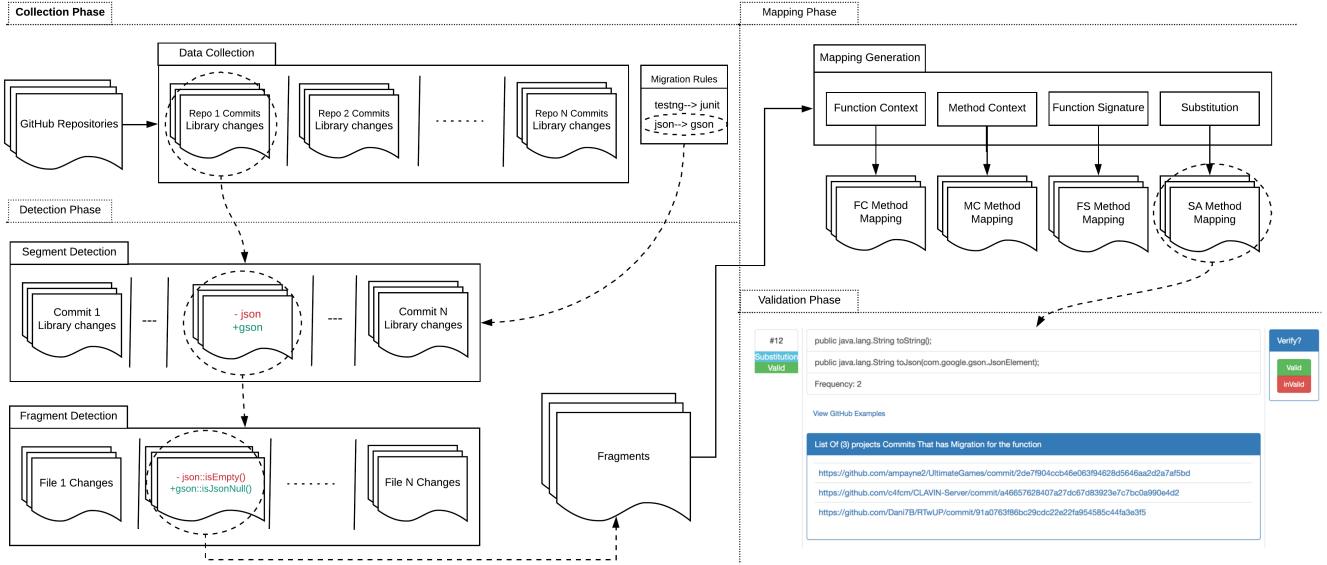

Figure 3: Approach Overview.

Indeed, we perform a straightforward static code analysis because a migration does not require the physical removal of the library to be retired from the project, as the retired library may still loaded in project through its *pom.xml* file; however none of the library's methods are used in the project's source. Once the segment end is located, we keep scanning previous commits in a backward fashion, looking for the beginning commit which contains the beginning of the fragment, *i.e.*, the first code change related to the replacement of any retired library method. After locating all segments for a given migration, it is important to keep track of source and target libraries versions for each segment to avoid backward incompatibility in case of an API change between two versions of the same library.

**Fragment Detection:** The fragment detection is responsible for the source code fragments related to the library migration changes as shown in Figure 3. It clones the project source files that are changed in the commits belonging to the identified time segments. We apply the Git's *Unified Diff Utility* command between the changed files to generate fragments, if any. A fragment is a continuous set of lines that have been changed along with contextual unchanged lines. Only fragments containing removed (resp., added) methods from the source (resp., target) library are considered valid. We retained a total of 8,938 fragments that we index in our dataset.

*C. Mapping Phase*

In this step, we generate method mappings from the identified fragments using our approach, which we label Substitution Algorithm (SA).

SA starts by sorting the identified code fragments using *Heap Sort* (cf. Algorithm 1). The sorting process is based on two attributes. **First**, the number of methods per code fragment which refers to the number of added and removed methods in a fragment. **Second**, the frequency of a fragment, *i.e.*, how many times a fragment appears across all projects.

If two code fragments have the same number of methods, then the fragment that has a higher frequency is moved before the fragment with lower frequency. The process moves the fragments that have less number of methods to the beginning of the list. Thereafter, SA iterates through all the identified fragments, starting from those with higher frequency, and searches for intersections $ISet \leftarrow fragment_1 \cap fragments_2$ between each fragment and the remaining set of fragments. Two fragments are considered to bear a not-null intersection $ISet$, if they share at least one common added and removed method. When an intersection exists, we remove shared methods $ISet$ from $fragment_1$ using $fragment_1 \leftarrow update(fragment_1 - ISet)$, and $fragment_2$ using $fragment_2 \leftarrow update(fragment_2 - ISet)$, and we add $ISet$ as new fragment $fragments \leftarrow add(ISet)$. Then, the algorithm iterates back to the sorting process. This process continues until there are no more intersections that can be found between all fragments in the list.

The intuition behind this process resides on using fragments with *one-to-one* and *one-to-many* mappings, which are more frequent to be seen in the identified fragments,

## Algorithm 1 Substitution Algorithm (SA)

**INPUT:** *fragments* - List of fragments, every fragment has list of added methods, list of removed methods.
**OUTPUT:** List of method mapping.

1: **procedure** SUBSTITUTION(*fragments*)
2:   *loop*:
3:     *fragments* ← HeapSort(*fragments*)
4:     **for all** $fragment_1 \in fragments$ **do**
5:       **for all** $fragment_2 \in fragments$ **do**
6:         $ISet \leftarrow fragment_1 \cap fragment_2$
7:         **if** *ISet* **then**
8:           $fragments \leftarrow update(fragment_1 - ISet)$
9:           $fragments \leftarrow update(fragment_2 - ISet)$
10:           $fragments \leftarrow add(ISet)$
11:           **goto** *loop*.
12:         **end if**
13:       **end for**
14:     **end for**
15:     *newFragment* ← LD(*fragments*)
16:     **if** *newFragment* **then**
17:       *fragments* ← *newFragment*
18:       **goto** *loop*
19:     **end if**
20:   **return** *fragments*
21: **end procedure**

**INPUT:** *fragments* - List of fragments have N-M method mapping.
**OUTPUT:** *newFragment* - Fragment has one added, and one removed method that have highest similarity score between method's description.

22: **procedure** LD(*fragments*)
23:   *maxScore* ← Average of similarity score
24:   **for all** $fragment \in fragments$ **do**
25:     **for all** $rmFun \in fragment$ **do**
26:       **for all** $addFun \in fragment$ **do**
27:         $score \leftarrow CSLD(rmFun_{Des}, addFun_{Des})$
28:         **if** *score* >= *maxScore* **then**
29:           *newFragment* ← *rmFun*, *addFun*
30:           *maxScore* ← *score*
31:         **end if**
32:       **end for**
33:     **end for**
34:   **end for**
35:   **return** *newFragment*.
36: **end procedure**

to reduce the cardinality of *many-to-many* fragments, by splitting based on any common *one-to-one* or *one-to-many* mappings. Sorting the fragments prior to applying the intersection gives the opportunity to split larger fragments using smaller, yet relevant, fragments instead of performing random intersections between fragments. Our approach is based on *Heap Sort* in this phase.

Once all intersections are completed, *LD(fragments)* iterates through fragments with *many-to-many* mappings, with the aim of splitting them further using the lexical similarity. For each fragment, our approach calculates the similarity score $CSLD(rmFun_{Des}, addFun_{Des})$, between the description of the removed methods $rmFun_{Des}$, and the description of added methods $addFun_{Des}$ by following two steps:

**First**, it extracts the key phrases[19] for $rmFun_{Des}$, and $addFun_{Des}$ to keep only relevant words. In our study we used the Microsoft Text Analytic API [5] to extract important key phrases from the text.

**Second**, calculate *Term Frequency–Inverse Document Frequency (TF-IDF)* for key phrases of $rmFun_{Des}$, and $addFun_{Des}$ that generate vector of numeric numbers for $rmFun_{Des}$, and $addFun_{Des}$, then apply *Cosine Similarity* between two vectors that is a measurement of how similar are two vectors based on the dot product of their magnitude [20]. If the similarity score is greater than or equal to the average similarity of the top detected *one-to-one* correct mappings, then SA creates a new fragment *newFragment* that contains these two methods and restarts the intersection process.

SA terminates the search and returns a list of fragments, when there are no more intersections to be found, and no more newly created mappings based on methods similarity. Each fragment contains a unique method mapping.

As an illustrative example of SA's workflow, we use the following migration rule *json → gson*, and four fragments 1, 2, 3 and 4, each fragment has a frequency of one as they have been extracted once during *detection phase* as it shows in Figure 4.

In iteration(A), the four fragments are sorted in ascending order using *HeapSort* based on the number of methods per-fragment and the frequency of fragment appearance.

Thereafter, during the intersection process, SA identifies possible intersection(s) between fragments 1, and 4, since both fragments share the same methods *get(int)* →*getAsLong()*. This intersection increases generates two fragments : fragment 1, which frequency increases by one and a new fragment 5 that are inserted in the current list of fragments, fragment 4 is now discarded.

In iteration(B), the current fragments are sorted again. During the search for intersections, another fragment is found between fragment 5, and fragment 2, as both contain the methods *toJSONString()* →*toString()*. This intersection generates new fragments 6, 7, and 8, while fragment 5 and 2 are discarded and the other fragments remain unchanged.

In iteration(C), fragments are sorted again. We observe that there are no new intersections found between the current fragments, therefore SA iterates through all *many-to-many* fragments with the aim of finding lower cardinality, interleaved mappings that can be extracted using lexical similarity between added and removed methods. In this case, only fragment 3 is a candidate. Library Documentation (*LD*) takes as input the methods' description from the API documentation to generate the key phrases [19]. Thereafter, it calculates the cosine similarity between the key phrases of all possible *one-to-one* combinations of the added and removed methods. *LD* returns the mapping with the highest similarity between the removed/added methods.

---
[5]https://goo.gl/exSkku

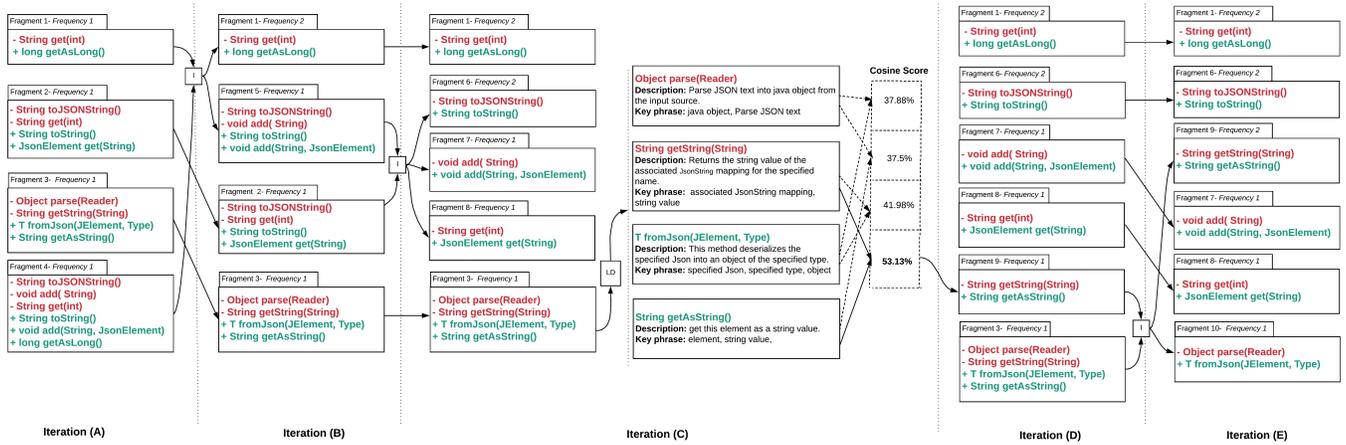

Figure 4: SA illustrative example using 4 fragments of migration between json and gson as input.

We observe from the example that the methods *getString(String)* →*getAsString()* have the highest similarity score, *i.e.*, 53.13%, which is close to the average cosine similarity of the already identified fragments 1, 6, 7, and 8. Thus, these two methods are considered as a new fragment that is added to the list.

In iteration(D), fragments are sorted again. Then, during the intersection process, SA identifies an intersection between fragments 9, and 3, since both fragments have *getString(String)* →*getAsString()*. This intersection increases the frequency of fragment 9 while creating a new fragment 10. while the other fragments remain unchanged.

In iteration(E), fragments are sorted again. Then, during the intersection process, SA cannot find any new intersection between the current fragments. Therefore, SA iterates through all *many-to-many* fragments. Since there are no *many-to-many* fragments, SA outputs each fragment as a final method mapping. In this example, the output of SA is a set of 6 mappings.

### D. Validation Phase

Most of method mapping that generated by SA already verified as valid or not valid method mapping by study[17]. For rest of method mapping that SA generated, we conducted a manual inspection process similarly to our previous study[17] by building a publicly available web portal[6] for the software engineering community that shows the list of library migration related-project commits for each method mapping. The authors then decide the correctness of the rule by manually checking the different method mappings in the list of commits, which constitute the ground truth. For example, from the project Selenium Grid Extras v1.1.9 [7], we observe a valid mapping between *put(key, description)* and *addProperty(key, description)*[8].

### VI. Experimental Design

We design our experimental study to mainly assess the accuracy of substitution algorithm (SA) to detect method mapping in compare with other approaches. We applied SA on same fragments that used by previous study[17] that compared three state-of-the-art approaches, namely Teyton et al. (FC) [5], Nguyen et al. (FS) [10], and Schäfer et al. (MC) [6]. We design our methodology to answer the two following research questions.

- **RQ1. (Accuracy)** To what extent SA is able to detect developer-performed method mappings?
- **RQ2. (Effectiveness)** How effective SA in detecting all the mappings with fewer fragments in comparison with existing approaches?

To answer RQ1, we evaluate the accuracy of our SA approach in detecting correct method mappings. We compare the SA mapping results with a ground truth set of the manually verified mappings, mined from 57,447 Java projects which provided by Allamanis et al. [21], extracted from migration rules that are manually validated and provided by Teyton et al. [22]. The accuracy of our approach is measured based on widely used metrics, TPR, and f-measure as follows:

---

[6]http://migrationlab.net/index.php?cf=icpc2019
[7]http://migrationlab.net/redirect.php?cf=icpc2019&p=2
[8]Line 75 in JsonResponseBuilder.java, in the following commit : http://migrationlab.net/redirect.php?cf=icpc2019&p=1

**True Positive Rate(TPR).** It denotes the ratio of correctly extracted method mappings by all expected mappings.

$$TPR(x) = \frac{Vx}{Ux}$$

where $Vx$ is the total number of valid mappings and $Ux$ is the total set of manually validated mappings.

**f-measure.** To measure which approach has better performance, we use f-measure as the weighted harmonic mean of both precision and recall.

$$f-measure(x) = \frac{2 * Precision * Recall}{Precision + Recall}$$

To answer RQ2, we measure the ability of our SA approach in generating the expected mappings with fewer code fragments, in comparison with the three considered state-of-the-art approaches. Instead of providing each approach with all the code fragments, across projects, they are gradually fed with randomly selected fragments while measuring their performance in terms of extracting all the correct mappings. To perform this experiment, we used the manually validated mappings to create synthetic fragments. The performance of each approaches under study is tested in three different settings as follows:

- **(A) One-to-one:** All the randomly created fragments contain a random number of *one-to-one* mappings.
- **(B) Many-to-many:** All the randomly created fragments contain a random number of *one-to-one*, *one-to-many*, and *many-to-many* mappings.
- **(C) Library Documentation:** All the randomly created fragments contain a random number of *one-to-one* mappings. Breaking larger fragments with library documentation between methods is enabled.

For each setting, we perform three experiments where the size of randomly created fragments is 5, 10 and 20. We generate a different number of fragments between 5-1500. We select these ranges to cover all possible real-word scenarios. To deal with the stochastic nature of the experimentation, we run each experiment instance 30 times; then we take the average f-measure for each approach.

## VII. Results

### A. Results for RQ1.

We calculate the TPR, and f-measure of the mappings generated by SA approach as well as FC [5], FS [10], and MC [6].

Table I shows the results achieved by each of the four approaches applied on the same dataset. On average, we observe that the accuracy score(f-measure) achieved by SA is higher than the three other approaches by 12%. The performance achieved by SA could be justified by the intersection process which aims at detecting low-frequency mappings that are not detected by FC, MC, and FS. In terms of TPR, SA achieved the best average TPR score (82.1%). That is mean that SA detects a larger number of correct mapping in compare with three state-of-art approaches.

In summary, the qualitative analysis of 9 migration rules has demonstrated that that SA's accuracy(f-measure) has an average of 75.2% while the maximum accuracy scored of the other approaches is 63.3%. Thus, SA increased the accuracy of the state-of-the-art by 12%.

### B. Results for RQ2

Figure 5 shows how each approach performs on a different number of methods and fragments.

We observe from the Figure that SA clearly requires much less number of code fragments than the three other approaches to reach an f-measure score of 100% in all settings, namely (A) *one-to-one*, (B) *many-to-many*, and (C) library documentation. We notice that the achieved f-measure stabilizes after achieving 100%, regardless of the number of fragments.

Furthermore, results in Figure 5 indicate that FC follows a similar convergence pattern as SA, but it requires more fragments to reach 100%. In addition, we note that the FC approach achieves a relatively less f-measure in (B) *many-to-many* in comparison with (A) *one-to-one*, for the same number of methods and fragments. For example, in Figure 5-(A) with up to 10 methods per fragment, FC has an f-measure of 93% per 21 fragments, while it has an f-measure of 88% per 21 fragments in Figure 5-(B) with up to 10 methods per fragment.

We also observe that the f-measure score of MC approach cannot reach 100% regardless of the number of fragments, because this approach can not detect multiple mapping in one fragment. Increasing the number of fragments, increase the possibility of having more *many-to-many* fragments for that reason, f-measure goes down when we increase the number of fragments.

Another interesting observation is that the nature of FS relies on the closeness of the naming practices followed by the different library developers, to find good matching between methods. Thus, increasing the number of fragments does not increase the performance of FS, instead, involving fragments with a larger set of methods increases the approach proneness to false positives, because it increases the probability of its inability to distinguish between methods when their signatures are similar. As shown in Figure 5, FS achieves an f-measure score of 69% in (A) *one-to-one* setting with 10 methods, then its f-measure score tends to decrease to 56%, when we increased the number of fragments to 1401 fragments. This observation applies to all experiment instances as shown in the figure. Moreover, we notice that the FS approach starts at a different f-measure score for different experiment instances, because it essentially depends on the similarity of the randomly selected mappings in the code fragments.

Table I: TPR and F-measure of the method mappings.

| Migration Rule | FC | | MC | | FS | | SA | |
|---|---|---|---|---|---|---|---|---|
| | TPR | f-measure | TPR | f-measure | TPR | f-measure | TPR | f-measure |
| logging→slf4j | 25% | 40% | 13% | 23% | 31% | 46% | 98% | 93% |
| easymock→mockito | 57% | 51% | 46% | 17% | 50% | 60.27% | 83% | 59.77% |
| testng→junit | 56% | 54% | 49% | 40% | 76% | 77% | 90% | 85% |
| slf4j→log4j | 78% | 70% | 73% | 73% | 57% | 64% | 100% | 90% |
| json→gson | 76% | 43% | 36% | 33% | 43% | 53.06% | 57% | 53.13% |
| json-simple→gson | 70% | 50% | 40% | 40% | 50% | 55.56% | 70% | 56% |
| Collection→guava | 78% | 68% | 78% | 81% | 73% | 77% | 78% | 83% |
| gson→jackson | 50% | 36% | 37% | 40% | 54% | 65% | 63% | 69.77% |
| sesame→rdf4j | 100% | 72% | 100% | 72% | 88% | 72% | 100% | 88% |
| Average | 65.5% | 53.8% | 52.4% | 46.5% | 58% | 63.3% | 82.1% | 75.2% |

## C. Discussions

In this section, we provide further discussions and insights about the obtained results.

**Unresolved method mappings:** Figure 5-(C) shows how different approaches perform when we enable the resolution of "unresolved" fragments that may have more than one valid mapping split using the similarity of method library documentation. We notice that SA is able to reach an f-measure of 100% with a relatively less number of fragments, in comparison with its performance without the use of library documentation similarity.

Overall, the achieved results indicate that SA is less prone to false positives. Even in the cases when SA is unable to split larger fragments, it recommends them as is. For example, from the experiment instance reported in Figure 6-(A), we found some "unresolved" fragments that are generated by SA, but with no false positive results. The fragment is a combination of two correct method mappings *toJSONString() →toString()*, and *get(int) →get(String)*. SA was unable to resolve these two fragments as it did not detect any fragment containing only either *toJSONString() →toString()* or *get(int) →get(String)*.

For the FC approach, the unresolved fragments tend to generate large number of false positives. For instance, as shown in Figure 6-(B), one of the unresolved fragments achieved by FC contains a method named *add(String, JsonElement)* which could be replaced by any of the removed methods. This is a false positive mapping generated by FC as these two removed methods have the same frequency with these three added methods. Having these cases lead library documentation to generate false positives as well and may not detect the correct method mapping, in case the false added methods which a method that added to the fragment during the solving fragments process or by code refactoring has a high similarity with one of the removed methods.

For the MC approach, the size of unresolved fragments is most likely large, because MC cannot resolve *many-to-many* mappings. This leads to increasing the chances of false positives when comparing the similarity between a larger set of methods descriptions. Indeed, the less is the number of methods in code fragments, the more accurate are the results. For this reason, we did not observe a significant effect of similarity in methods descriptions for the MC approach. The methods description similarity can either increase or decrease the mappings accuracy for MC. For example, for ten methods in Figure 5-(B, and C), MC achieved an f-measure of 5.2% in (B), when the number of the fragments reaches 101. While MC's f-measure is 12.8% in (C), when the number of the fragments reaches 101. This indicates that the methods description similarity increases the coverage.

FS does not generate *many-to-many* unresolved fragments because it solves *one-to-one* or *one-to-many* method mappings only. Therefore, the methods description similarity does not have large effect on the approach's accuracy.

Table II: Number of library documentation TPR.

| Number of methods | Times use library documentation | | | |
|---|---|---|---|---|
| | FC | MC | FS | SA |
| 2 | 89 | 184 | 0 | 0 |
| 5 | 497 | 667 | 0 | 76 |
| 10 | 2244 | 2538 | 0 | 76 |
| 20 | 4035 | 4433 | 0 | 67 |

Different approaches require a different number of times that approach use library documentation to detect method mapping from "unresolved" Fragments to reach the best TPR score. Table II shows that MC requires a large number of method calls in comparison with the three other approaches since it cannot resolve *many-to-many* mappings. We need to apply *LD* to address this limitation. While FC requires fewer method calls than MC, and greater than SA. While SA requires the minimum number of using library documentation to detect method mapping from "unresolved" Fragments in comparison with FC, and MC.

In summary, SA requires fewer fragments than existing approaches. Furthermore, the library documentation helps SA to reach 100% of f-measure score earlier. Therefore, this answers RQ2 indicating that SA is effective in detecting all the mappings, when compared to the three other approaches.

**Positive outcomes:** In scenarios where there is a sufficient fragments, both SA and FC can reach an accuracy score of 100%. However, SA is considered better as it requires less number of fragments to reach 100% accuracy.

**Negative outcomes:** On the other hand, an increasing number of fragments makes the accuracy of MC and FS worse. This could be mainly due to two different reasons. First, FS is able to map a limited number of methods that have a similar signature, so it does not rely on counting the number of detected mappings in the fragments. Therefore, increasing the number of fragments will not help in finding more mappings, instead of leading to more false/positives mappings. Second, MC maps blocks of code, thus increasing the number of fragments will increase the possibility of having segments with *many-to-many* methods, and which leads to more false positives.

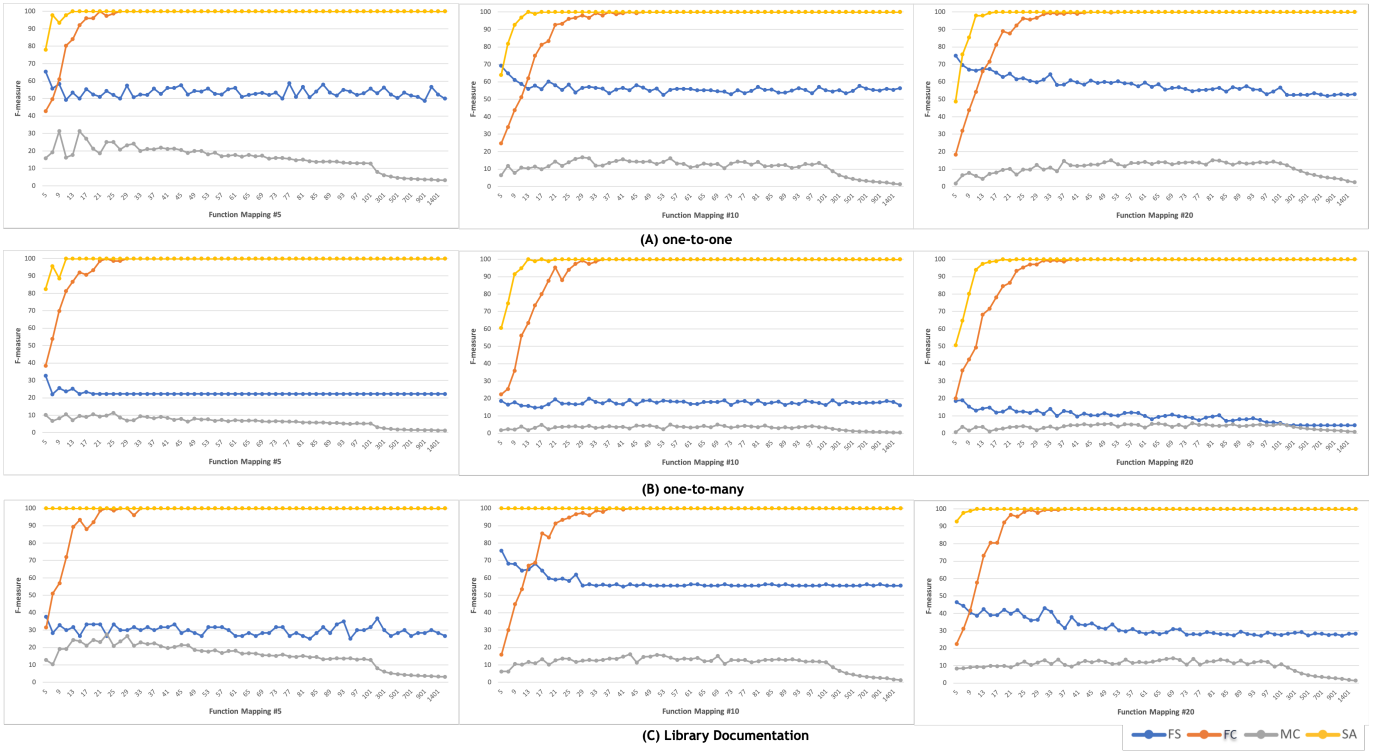

Figure 5: f-measure using randomly created 5-1500 fragments for 5, 10, and 20 method mapping, over 30 runs.

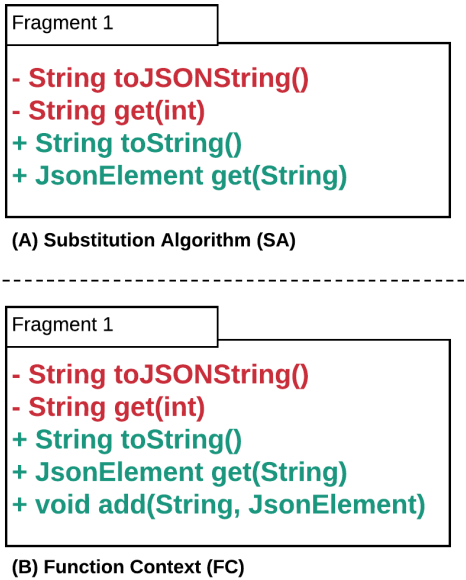

Figure 6: Example of "unresolved" Fragments.

In summary, it is not clear what is the best number of fragments that we should reach to reach the best accuracy for FS and MC. However, for SA and FC, it is intuitive that increasing the number of fragments helps in reaching a better accuracy.

## VIII. Conclusion

This study addressed the problem of mining developer decision in migrating third-party libraries. we have described a novel approach that detects all the method mappings performed by developers when migrating between two different libraries. Our approach combines the mined method change patterns with method related documentation similarity to accurately detect mappings between removed and added methods. We evaluated our approach by mining the method-level changes of 9 popular library migrations across several open-source Java projects. The qualitative and comparative analysis of our experiments indicates that our approach significantly increases the accuracy of detecting manually-performed method-level mappings by an average accuracy of 12%, with fewer required fragments, in comparison with existing state-of-the-art studies.